\documentclass[aip, twocolumn, reprint]{revtex4-1}

\usepackage{amssymb,amsmath,xcolor,graphicx,xspace,colortbl,rotating} %
\usepackage[raggedrightboxes]{ragged2e} 
\usepackage{amsfonts}  
\usepackage{amsmath}  
\usepackage{amssymb}  
\usepackage{boxedminipage}  
\usepackage{float}  
\usepackage{graphics}  
\usepackage{graphicx}  
\usepackage{ragged2e}  
\usepackage{wrapfig}  
\usepackage{xcolor}  
\graphicspath{{NRnsswitching_graphics/}{NRnsswitching_tcache/}{NRnsswitching_gcache/}}
\DeclareGraphicsExtensions{.pdf,.eps,.ps,.png,.jpg,.jpeg}
\providecommand{\U}[1]{\protect\rule{.1in}{.1in}}

\begin{document}
\title{Digital electric field induced switching of plasmonic nanorods using an electro-optic fluid fiber
}
\author{Sebasti{\'a}n Etcheverry}

\affiliation{Department of Applied Physics, KTH, 10691 Stockholm, Sweden}
\affiliation{Department of Fiber Optics, RISE Acreo, Box 1070, SE-164 25, Kista, Sweden
}

\author{Leonardo F. Araujo}

\affiliation{Department of Physics, Pontif{\'\i}cia Universidade Cat{\'o}lica do Rio de Janeiro (PUC-RIO), Rio de Janeiro, 22451-900, Brazil }

\author{Isabel C. S. Carvalho}

\affiliation{Department of Physics, Pontif{\'\i}cia Universidade Cat{\'o}lica do Rio de Janeiro (PUC-RIO), Rio de Janeiro, 22451-900, Brazil }

\author{Walter Margulis}
\email{walter.margulis@ri.se
}

\affiliation{Department of Applied Physics, KTH, 10691 Stockholm, Sweden}
\affiliation{Department of Fiber Optics, RISE Acreo, Box 1070, SE-164 25, Kista, Sweden
}

\author{Jake Fontana}

\email{jake.fontana@nrl.navy.mil}

\affiliation{Naval Research Laboratory,~4555 Overlook Ave. Washington, D.C. 20375, U.S.A.}

\begin{abstract}We demonstrate the digital electric field induced switching of plasmonic nanorods between ``$1$'' and ``$0$'' orthogonal aligned states using an electro-optic fluid fiber component.  We show by digitally switching the nanorods, that thermal rotational diffusion of the nanorods can be circumvented, demonstrating an approach to achieve submicrosecond switching times.  We also show, from an initial unaligned state, that the nanorods can be aligned into the applied electric field direction in $110$ nanoseconds. The high-speed digital switching of plasmonic nanorods integrated into an all-fiber optical component may provide opportunities for remote sensing and signaling applications.
\end{abstract}

\volumeyear{year}
\volumenumber{number}
\issuenumber{number}
\eid{identifier}
\date{date}
\received[Received text]{date}
\revised[Revised text]{date}
\accepted[Accepted text]{date}
\published[Published text]{date}
\startpage{101}
\endpage{102}
\maketitle
\bigskip In general, the permanent or induced dipole moment and resulting polarizability of a molecule is too small to couple to external electric fields to overcome disordering thermal forces, preventing alignment. If anisotropic molecules are condensed into a liquid crystal phase, then the additional van der Waal forces from the near-neighbor interactions increases the polarizability to enable alignment of the molecules and control the optical properties.  The electric field induced alignment of anisotropic molecules in liquid crystal phases has enabled disruptive technologies such as smart phones and flat screen displays.\cite{11,22}

The switching time of these materials depends on the sum of their \textit{on}- and \textit{off}-times.  The \textit{on}-time needed to align the molecules into the direction of the applied electric field is predominately set by the magnitude of the field applied, $\tau _{on} \approx \gamma /\varepsilon E^{2}$, where $\gamma $ is the viscosity, $\varepsilon $ is the dielectric permittivity and $E$ is the electric field.  The \textit{off}-time is related to the thermal rotational diffusion of the liquid crystal molecules and typically is the limiting factor to determine the overall switching time.  In the case of liquid crystals, the near-neighbor interactions create strong electrohydrodynamic coupling, leading to a slow characteristic \textit{off}-time, $\tau _{off} \approx \gamma d^{2}/K \approx ms$, where $d$ is the cell thickness and $K$ is the elastic constant of the liquid crystal. This well-known limitation has constrained potential electro-optic applications for decades. 

A recent elegant approach avoided re-alignment of the molecules altogether by rapidly electrically inducing a change in the refractive index of the molecules in a liquid crystal phase.\cite{33}  The response time of this system was $10^{ \prime }s$ of nanoseconds, yet the change in the optical properties was small and cannot be maintained for long times before the usual electric field induced alignment of the molecules occurs.

The electric field induced alignment of plasmonic nanorods is a paradigm to anisotropic molecules in liquid crystal phases.\cite{44}  A key advantage of plasmonic nanorods is that the polarizibility of a single nanorod in a dilute suspension is adequately large to couple to an external electric field, enabling alignment and the ability to tune the optical properties at visible and near infrared wavelengths.\cite{44,55} A significant consequence of this electro-optic mechanism is that the \textit{off}-time decreases by $1 ,000 -$fold ($\gamma L^{3}/k_{b}T \approx \mu s$,  where $L$ is the length of the nanorod, $k_{b}$ is the Boltzmann constant and $T$ is the temperature) compared to liquid crystals, due to the absence of near-neighbor interactions.\cite{66}  The electric field induced alignment of plasmonic nanorods alleviates the long-standing switching time limitation of liquid crystal based devices, potentially ushering in innovative display, filter and spatial light modulators technologies.\cite{66}

The switching time for the plasmonic nanorods, is limited to the order of microseconds, due to the thermal rotational diffusion of the nanorods when the field is switched \textit{off}. To break-through this limit, consider that the \textit{on}-time for the nanorods can become arbitrarily small if the applied electric field is large. Therefore if the nanorods are digitally switched between two orthogonal fields, holding the nanorods in a continuous ``$1$'' or ``$0$'' aligned state, then thermal relaxation can be circumvented, potentially leading to submicrosecond switching times.\cite{77}

\begin{figure}\centering 
\setlength\fboxrule{0.01in}\setlength\fboxsep{0.1in}\fcolorbox[HTML]{FFFFFF}{FFFFFF}{\includegraphics[ width=3.25in, height=1.87721021611002in,]{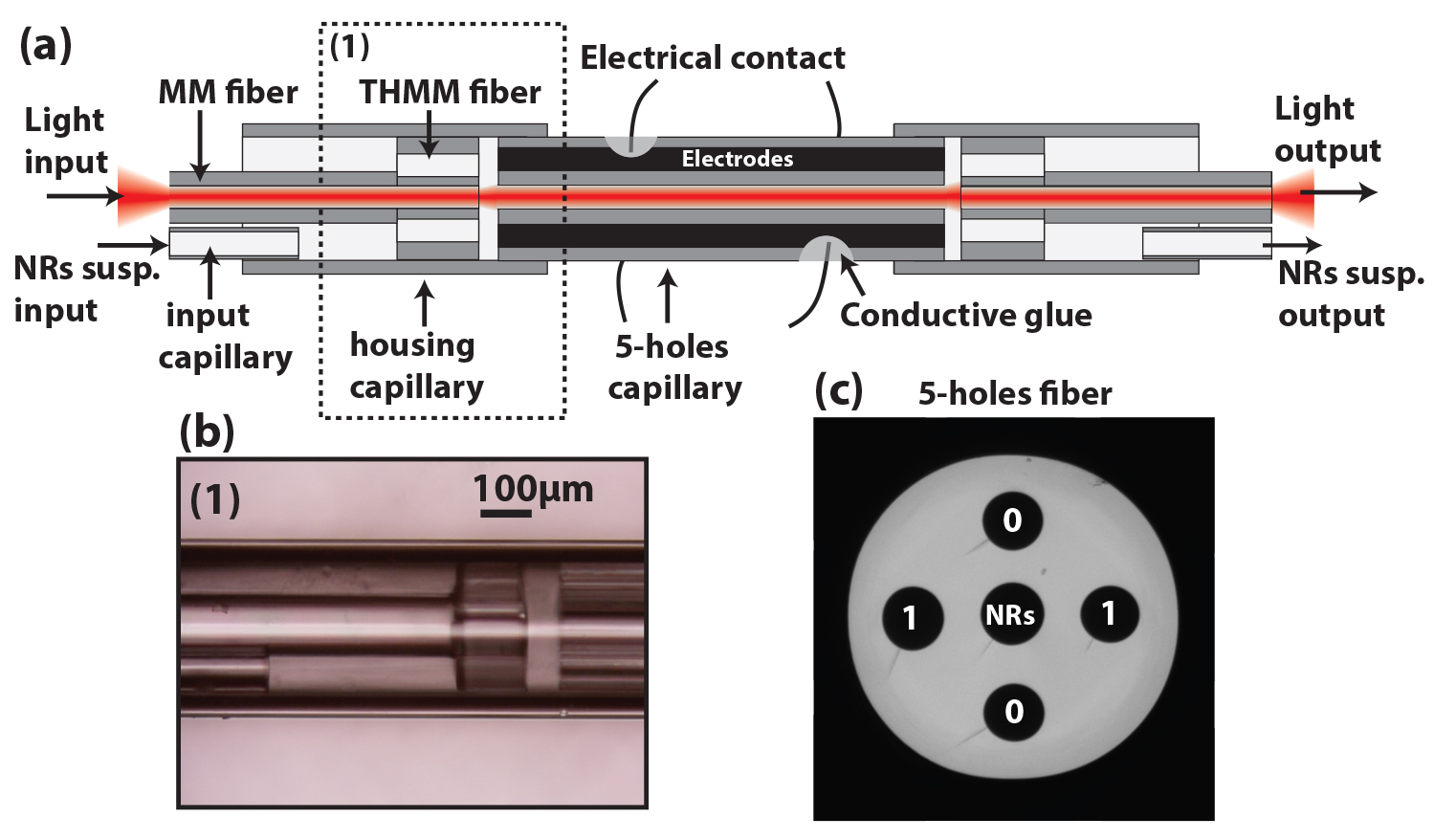}
}
\caption{ (a) Schematic of the fiber component. (b) Microscope image of the part (1) of the fiber component. (c) Microscope image of the 5-hole fiber.
}\label{1}\end{figure}

To accomplish this, gold nanorods (length/diameter=$75$/$25\ nm$) were synthesized using a wet-seed mediated method, coated with ligands (thiol terminated polystyrene, $M_{n} =5 ,000$ (Polymer Source, Inc.)) and suspended in toluene ($ \sim 10^{ -4}\ v/v \%$) to enable alignment and waveguiding inside the fiber.\cite{44,66,88}

The suspensions were placed into an electro-optic fluid fiber component, Fig.1(a).  Light was coupled from the left along a $62.5/125$ $\mu m$ (inner/outer diameter) multimode fiber (MM fiber) spliced to a $62.5/250$ $\mu m$ fiber which has two $50 -\mu m$ holes along its cladding (THMM, Fig. 1(a)).\cite{66}  The splice was fitted into a $250/330$ $\mu m$ capillary (housing capillary). A $90/125$ $\mu m$ capillary (input capillary), tapered to $ \sim 50/70$ $\mu m$, incorporates the nanorod suspension into the component via the housing capillary.  The MM fiber, input capillary and the housing capillary were sealed together with UV curing glue.

\begin{figure}\centering 
\setlength\fboxrule{0.01in}\setlength\fboxsep{0.1in}\fcolorbox[HTML]{FFFFFF}{FFFFFF}{\includegraphics[ width=3.25in, height=2.5971512052593133in,]{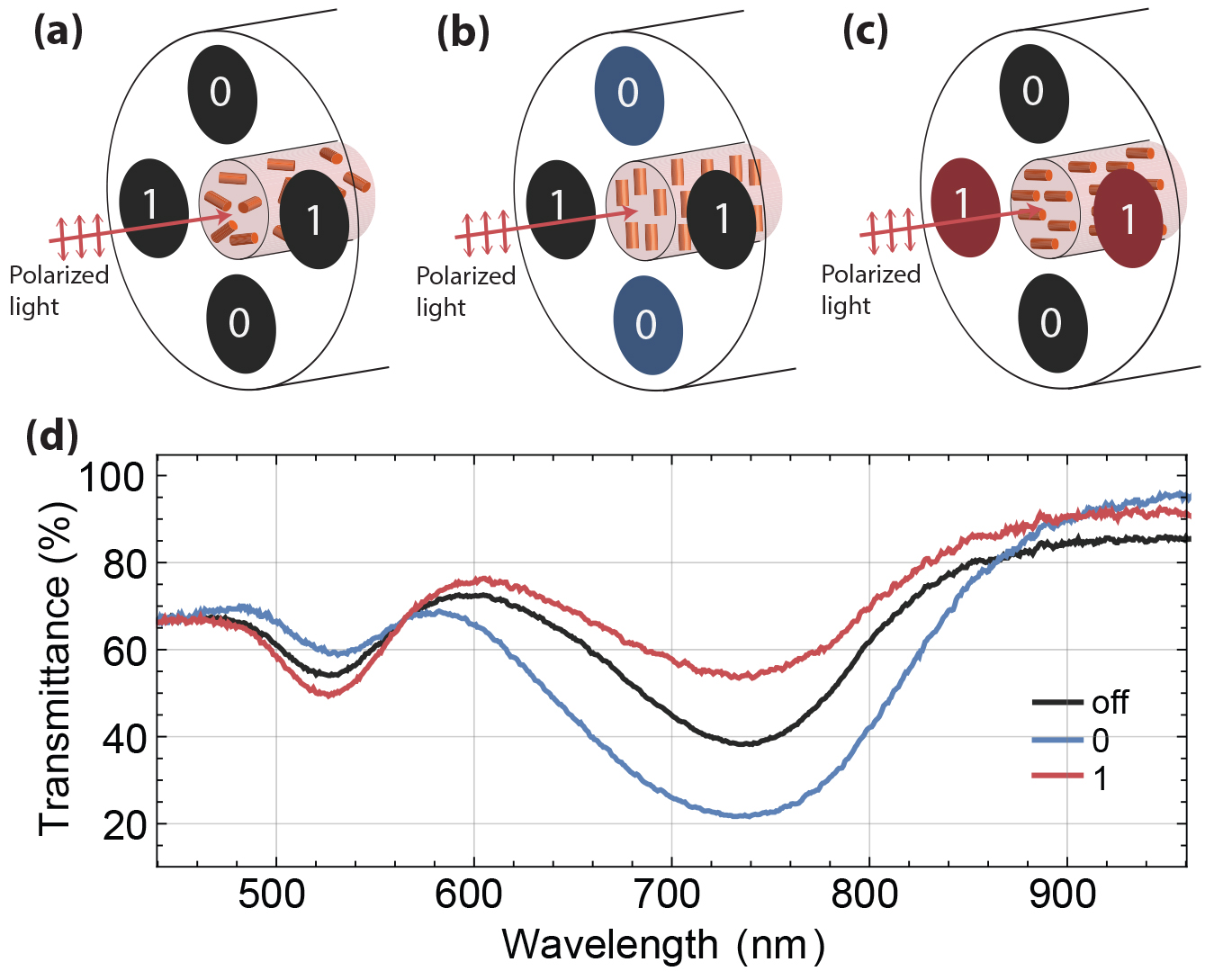}
}
\caption{Schematic of the 5-hole fiber without an electric field applied (a) and with the AC field applied for the $0$ (b) and $1$ (c) digital alignment states. The probe light was polarized along the $0 -$state direction.  The measurements were integrated for $100$$ms$. (d) Transmittance spectra collected from the component output.}\label{2}\end{figure}

The active fiber section consists of a $10$ $cm$ long 5-holes fiber, Fig. 1(c), having the $4$ outer $50\ \mu m$  diameter holes filled with BiSn,\cite{99}  and is placed on the right side of the housing capillary.The BiSn filled holes serve as orthogonal digital pairs of electrodes (channel $0$ and channel $1$). The electrical contacts for the electrodes are made by side-polishing the 5-hole fiber and gluing a copper wire to each electrode with silver based conductive glue.   A small gap ($ \sim 65\ \mu m$) from the THMM fiber end allows the pressurized $( \sim 10^{5}\ Pa)$ nanorod suspension to flow from the input capillary and into the central hole of the 5-holes fiber, where the nanorod interact with the applied electric fields.  The construction of the electro-optic fluid fiber component is symmetric after the 5-hole fiber allowing the nanorod suspension and transmitted light to be separated and collected.

The time-averaged spectral evolution of the nanorods as a function of applied electric field was measured by coupling linearly polarized white light (halogen lamp) into the component.  The output light from the component was coupled into a spectrometer (OceanOptics QE65000).  In order to have stable alignment, the electric field is an alternating field at $200$ $kHz$ and typically $2\ kV$.  The AC field prevents translation of the nanorods towards the electrodes and attachment to the inner walls.  There were small deviations of the distances between the electrodes for each channel due to fluctuation in the manufacturing process, Fig.1(c).  If a potential of $2\ kV$ was applied to each channel, then the electric field in the fiber core was calculated using COMSOL Mulitphysics $5.2$ to be $14.8\ V/\mu m$ for channel $0$  and $13.5\ V/\mu m$  for channel $1$.

Without an electric field applied to either channel, Fig. 2(a), the distribution of nanorods inside the central core of the 5-hole fiber is random with the transmittance shown in Fig. 2(d), black curve.  The randomly distributed nanorods have two absorption peaks leading to a decrease in the transmittance, one occurring at $740$ $nm$ from the long axis of the nanorod and the other at $530$ $nm$ corresponding to the short axis.

Fig. 2 demonstrates the nanorods can be aligned in the fiber component parallel ($0$), Fig. 2(b), and perpendicular ($1$), Fig. 2(c), to the probe light polarization.\cite{44}  When channel $0$ is switched \textit{on}, Fig. 2(d) blue curve, applying an AC signal, the nanorods align parallel to the field.  The probe light is polarized parallel to the applied electric field (channel $0$) therefore the transmittance decreases at $740$ $nm$  due to the increased absorption from the long axis of the aligned nanorods. Conversely, the transmittance increases at $530$ $nm$ from the decreased absorption from the short axis of the nanorods.  If channel 0 is turned \textit{off} and channel 1 is switched \textit{on}, Fig. 2(d) red curve, then the transmittance increases at $740$ $nm$ and decreases at $530$ $nm$ due to the polarization of the probe light being perpendicular to channel $1$.

\begin{figure}\centering 
\setlength\fboxrule{0.01in}\setlength\fboxsep{0.1in}\fcolorbox[HTML]{FFFFFF}{FFFFFF}{\includegraphics[ width=3.25in, height=2.2011225444340505in,]{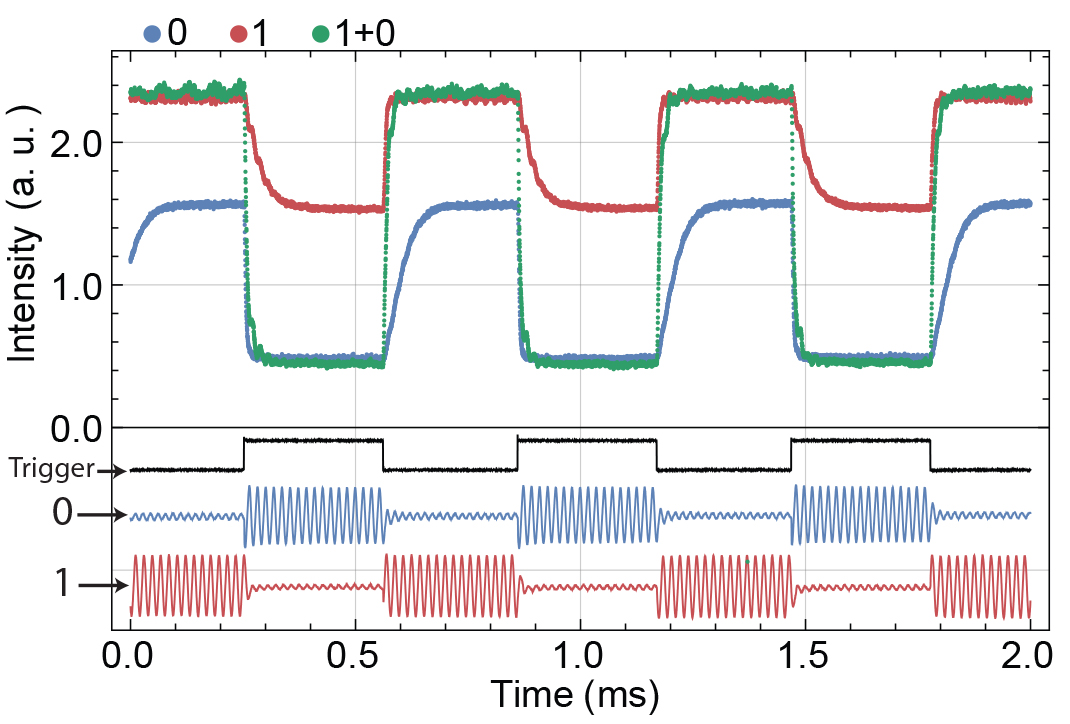}
}
\caption{Time-resolved digital nanorod switching. Intensity evolution for channel $0 -$blue, $1 -$red and green $1 +0$.  The high voltage trigger is shown in black and corresponding driving signals for each channel below, $\omega  =200\ kHz$,  $E =14.8\ V/\mu m$ (channel $0$) and $13.5\ V/\mu m$ (channel $1$).}\label{3}\end{figure}The time-resolved switching of the nanorods is reported in Fig. 3.  The white light source was replaced with a diode laser (Thorlabs HL7302MG).  The laser light wavelength was $730$ $nm$, corresponding to the absorption peak from the long axis of the nanorods.  The light was linearly polarized parallel to the $0$  channel.  A fast silicon detector (Thorlabs PDA10A) was used to measure the intensity of the light output from the component. Two synchronized high voltage (HV) switches, based on HV power mosfets (IXYS 4500V), were used to toggle the driving signal in each channel ($0 -$blue/$1 -$red) \textit{on} and \textit{off}.  The channels were triggered with a $0.6$ $ms$ period square wave, Fig. 3 (black curve), channel $0$ when the trigger is high and channel $1$ when the trigger is low.

The blue curve in Fig. 3 shows the response as just channel $0$ is switched \textit{on} and \textit{off}.  With the channel \textit{on} the nanorods quickly align, leading to a decrease in the transmittance at $730$ $nm$.  When channel $0$ is switched \textit{off}, the nanorods begin to randomize due to the thermally mediated rotational diffusion, increasing the transmittance. In contrast, if just channel $1$ is switched \textit{on}, then the transmittance rapidly increases (red curve) and decays to the randomized intensity when channel $1$ switches \textit{off}.  The period of the driving signal $\left (5\ \mu s\right )$ was set to be faster than the thermal relaxation time of the nanorods, therefore when the nanorods are exposed to the electric field they do not have time to thermally decay per AC cycle.

If the driving signal is toggled between channels $0$ and $1 ,$ then Fig. 3 (green cruve) clearly shows the digital response of the nanorods, switching between the $0$ and $1$ aligned states, thereby demonstrating the removal of the thermal rotation diffusion constraint from the switching mechanism. By fitting an exponential function to the leading or falling edge of the signals and retrieving the $1/e$ values, the \textit{on- }and\textit{ off}-times are defined. The digital \textit{on}- and \textit{off}-times in Fig. 3 (green curve) are both $8$ $\mu s$, at least three times faster than thermal mediated diffusion,\cite{66} and is limited only by the internal response time of the HV switch and driving frequency. The modulation depth for the digital response is the sum of the individual channels, thereby doubling the contrast ratio for the transmitted signal at $730\ nm$.

Since the digital switching time is limited by the speed of the electronics in the experiment. The response time was characterized by measuring the time to switch from a random to aligned state using a single pair of electrodes and a Behlke GHTS 100 push--pull switch for nearly rectangular pulse excitation. The $3$ $\mu s$ duration rectangular pulse (black curve) with $E =32.6\ V/\mu m\ $(channel $0$) $29.7\ V/\mu m$  (channel $1$) is applied to the component in Fig. 4(a).  The nanorods quickly respond to the presence of the electric field, yielding \textit{on}-times as fast as $110\ ns$ (channel $0$) and $160\ ns$ (channel $1$). As the pulse shuts \textit{off} and the electric field is removed, the nanorods then begin to thermally randomize, leading to characteristic \textit{off}-times of $33\ \mu s$ (channel $0$) and $22\ \mu s$ (channel $1$).\cite{66}  The differences for the switching times between both channels may have various contributions; these include small variations in the core and electrode separations, leading to differences in the electric fields for each channel, the creation of a (ionic) surface charge layer on the walls of the central hole during the application of the voltage pulse and nonlinear response and saturation of the photodetector used.  For the \textit{off}-times all of these effects contribute to a deviation from a single exponential decay of the signals in Fig. 4(a). If the \textit{off}-time signals are analyzed at intermediary times $\left (40 -70\ \mu s\right )$, then the aforementioned effects are eliminated and both channels follow the same exponential decay with an \textit{off}-time of $25\ \mu s$, Fig. 4(b).

\begin{figure}\centering 
\setlength\fboxrule{0.01in}\setlength\fboxsep{0.1in}\fcolorbox[HTML]{FFFFFF}{FFFFFF}{\includegraphics[ width=3.3in, height=3.22823086574655in,]{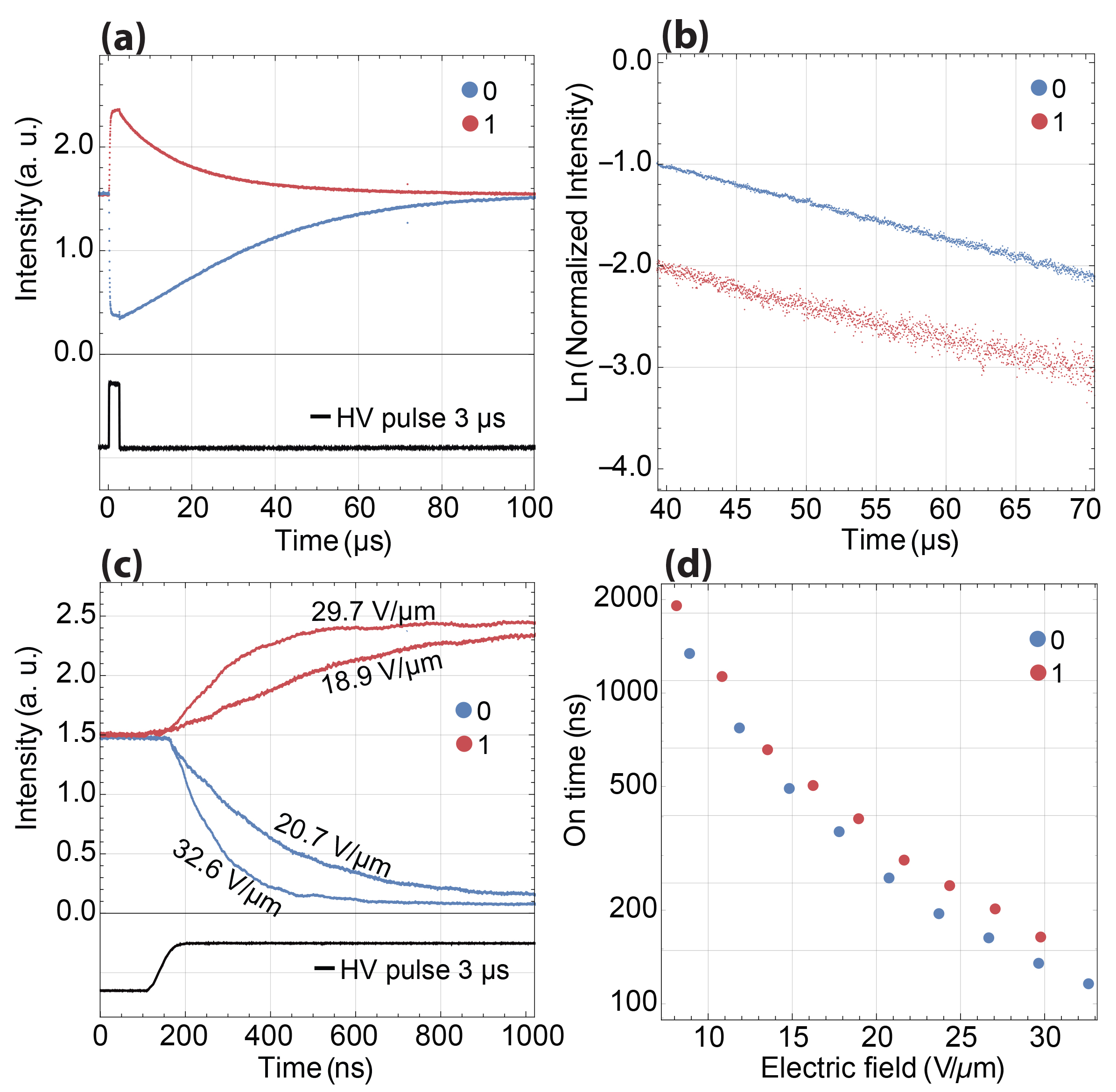}
}
\caption{ (a) Transmitted intensity through the fiber as the nanorods are aligned parallel ($0$) and perpendicular ($1$) to the probe light polarization using a $3\ {\mu}s$ rectangular pulse, $E =32.6\ V/\mu m$ (channel $0$) $29.7\ V/\mu m$ (channel $1$).  (b) Semi-log plot of the normalized intensities in (a) for intermediary times.  (c) Transmitted intensity for both channel as a function of electric field magnitudes.  (d) O\textit{n}-time versus electric field magnitude.
}\label{4}\end{figure}Fig. 4(c) and Fig. 4(d) are the responses, for both channels, as the electric field magnitude is varied, significantly influencing the \textit{on}-times.  The response is limited only by the electronic circuit, not a physical barrier from the electric field and nanorod coupling.  Therefore the \textit{on}-time could be improved in future experiments using enhanced circuitry.  These results indicate that digital switching between $0$ and $1$ states could be achieved at nanosecond time scales by developing advanced electronics for high voltage $\left ( >3\ kV\right )$ and high frequency $\left ( >100\ MHz\right )$.

The \textit{on}-\textit{off} ratio of this system can be further optimized. From a theoretical standpoint, all of the light polarized parallel to the alignment direction, at a wavelength matching the absorption peak associated with the nanorod's long-axis, can be absorbed if the density of nanorods and path length are sufficient. Therefore, the extinction ratio can in principle be infinite. In practice, the fibers used in this study are not polarization maintaining and some degree of ellipticity could limit the \textit{on-off} ratio measured. This could be improved using polarization maintaining fibers.

The applied voltage required to align the nanorods can be decreased through further engineering. For example, the hollow-core diameter and the mutual distance between the holes containing the electrodes can be smaller so that the same electric field can be achieved with a reduced voltage.  It is also possible to decrease the applied voltage by increasing the length, or volume, of the nanorods, which decreases the critical electric field needed to align the nanorods.\cite{44}

These electro-optic fluid fiber components could be potentially incorporated into existing optical fiber networks and used to modulate optical signals, such as a $GHz$ optical shutter or spatial light modulator.  The components may also be used to remotely sense large electric fields, for example in high-voltage transformer environments. Additionally, the components may be miniaturized and incorporated into microfluidic devices.

In summary, we developed an electro-optic fluid fiber that combines light, nanorod suspensions and electric fields into a single component.  Using this component, we showed the capability to digitally switch the plasmonic nanorods between two orthogonal aligned states using electric fields, demonstrating the removal of the thermal rotation diffusion constraint from the switching mechanism.  We also showed the nanorods can align into the applied electric field direction in $110$ $ns$, limited only by the experimental electronics.  These results may lead to exciting opportunities for the point-to-point delivery and modulation of light.

This work was supported by the Office of Naval Research Global (ONRG-NICOP-N62909-15-1-N016).  S.E. thanks Comisi{\'o}n Nacional de Investigaci{\'o}n Cient{\'\i}fica y Tecnol{\'o}gica (CONICYT), L.A. and I.C. thank Conselho Nacional de Desenvolvimento Cient{\'\i}fico e Tecnol{\'o}gico (CNPq) and Coordena{\c c}{\~a}o de Aperfei{\c c}oamento de Pessoal de N{\'\i}vel Superior (CAPES), Swedish Foundation for International Cooperation in Research and Higher Education (STINT) (006/12), W.M. thanks Swedish Research Council (VR), ADOPT Linnaeus Center in Advanced Optics and Photonics and STINT/CAPES, and J.F. thanks the Office of Naval Research for support.  S.E. and W.M. also thank Joao Pereira, Oleksandr Tarasenko and Leif Kjellberg for experiment help as well as Fredrik Laurell for support.

\ \bibliographystyle{apsrev4-1}
\bibliography{}
\end{document}